\documentclass{amsart}

\usepackage{amsmath,amssymb,amsfonts,amscd,amsthm,amsopn,epsfig}  
\usepackage{amsthm}
\usepackage{amssymb}
\usepackage{latexsym,verbatim}
\usepackage{mathrsfs}
\usepackage{enumerate}

\usepackage[utf8]{inputenc}

\usepackage{tikz}\usetikzlibrary{backgrounds,matrix,arrows,shapes,decorations.markings}
\newcommand{\btp}{\begin{tikzpicture}[baseline=-5pt,scale=0.25,line width=0.7pt]}
\newcommand{\etp}{\end{tikzpicture}}

\pgfdeclarelayer{background}
\pgfdeclarelayer{foreground}
\pgfsetlayers{background,main,foreground}

\usepackage{hyperref}
\hypersetup{
    citecolor=blue,        
}
\usepackage{caption}
\usepackage{float}

\newcommand{\overbar}[1]{\mkern 1.5mu\overline{\mkern-1.5mu#1\mkern-1.5mu}\mkern 1.5mu}

\usepackage{color}

\newcommand{\tr}{\text{Tr}}

\newcommand{\beq}{\begin{equation}}
\newcommand{\eeq}{\end{equation}}
\newcommand{\ben}{\begin{eqnarray}}
\newcommand{\een}{\end{eqnarray}}

\newcommand{\cH}{\mathcal H}

\newcommand{\mA}{\mathscr{A}}
\newcommand{\mB}{\mathscr{B}}
\newcommand{\mC}{\mathscr{C}}
\newcommand{\mD}{\mathscr{D}}

\def\y2{$Y(\mathfrak{sl}_2)$}

\newcommand{\so}{\scriptscriptstyle \rm I}
\newcommand{\st}{\scriptscriptstyle \rm I\hspace{-1pt}I}

\begin{document}


\title{Slavnov and Gaudin-Korepin formulas for models without $U(1)$ symmetry: the XXX chain on the segment}
\author{S.~Belliard}
\address{Laboratoire de Physique Th\'eorique
et Mod\'elisation (CNRS UMR 8089), Universit\'e de Cergy-Pontoise, F-95302 Cergy-Pontoise, France }
\address{Institut de Physique Th\'eorique, DSM, CEA, URA2306 CNRS Saclay, F-91191 Gif-sur-Yvette, France}
\email{samuel.belliard@u-cergy.fr}
\author{R.A.~Pimenta}
\address{Departamento de F\'{\i}sica, Universidade Federal de S\~ao Carlos,
Caixa Postal 676, CEP 13565-905, S\~ao Carlos, Brasil}
\address{
Physics Department, University of Miami, P.O. Box 248046, FL 33124, Coral Gables, USA}
\email{pimenta@df.ufscar.br}

\begin{abstract}
We consider the isotropic spin$-\frac{1}{2}$ Heisenberg chain with the most general integrable boundaries.
The scalar product between the on-shell Bethe vector and its off-shell dual, obtained by means of the modified algebraic Bethe ansatz,
is given by a modified Slavnov formula. The corresponding Gaudin-Korepin formula, \textit{i.e.}, the square of the norm, is also obtained.
\end{abstract}

\maketitle

\paragraph{\textbf{Introduction}}The algebraic Bethe ansatz (ABA) \cite{SFT,Skly88} is a powerful technique to study the spectral problem
of quantum integrable models, as well as to construct their correlation functions and compute physical
quantities \cite{KBI}. The possibility of expressing scalar products in a compact form \cite{Gau,GauMcCoyWu,K,Sla89} is a crucial
aspect of the method. Nevertheless, the application of the usual ABA to obtain the spectrum and
the eigenvectors of quantum integrable models becomes problematic for
some important cases, in particular in the presence of the non-diagonal boundaries. In the case of
the Heisenberg spin chain on the segment, this is a consequence of the
breaking of the $U(1)$ symmetry by off-diagonal boundaries. Many approaches have been developed to handle this problem, including
generalizations of the Bethe ansatz to consider special non-diagonal boundaries, see for instance \cite{gauge,TQ1,BCR12,PL13,AMS} and references therein, the SoV method \cite{FSW08,FGSW11,Nic12,FKN13,KMNT15}, the functional method \cite{G8},
the q-Onsager approach \cite{BK} and the non-polynomial solution from the homogeneous Baxter
T-Q relation \cite{LP14}.

\vspace{0.1cm}

Recently, the ABA has been generalized to include models with general boundary couplings \cite{BC13,B15,Cra14,BelPim15,ABGP15}. The
modified algebraic Bethe ansatz (MABA) has a distinct feature: the creation operator used to construct the eigenstates has an
off-shell structure which leads to an inhomogeneous term in the eigenvalues and in the Bethe equations of the model. We remember that such inhomogeneous term was firstly proposed
in the context of the off-diagonal Bethe ansatz (ODBA) method (see \cite{ODBAbook} for a review) and
recovered in the separation of variables (SoV) framework \cite{KMN14}. Let us also remark that the SoV basis \cite{FKN13} has been used to prove the off-shell equation for the creation
operator in the MABA context \cite{ABGP15} as well as to obtain the on-shell Bethe vector in the ODBA method \cite{retrieve1,retrieve2}.

\vspace{0.1cm}

Once the spectral level is understood, the next natural step is to consider the evaluation of scalar products
between Bethe vectors obtained in the MABA framework.
The calculation of scalar products within this context is a primordial step
to access the physical behavior of systems with general integrable open boundaries,
since it paves the way to consider form factors and correlation functions. This task has been recently initiated in the case of the twisted XXX spin chain \cite{BelPimtwist}, a
prototype model that can be described by the MABA. Modified Slavnov and Gaudin-Korepin formulas,
\textit{i.e.}, compact expressions
for the scalar product between an on-shell and an off-shell Bethe state and the square of the norm, have been conjectured.
We propose in this note similar formulas for the XXX spin chain on the segment. They are given in terms of a determinant involving
the inhomogeneous eigenvalue expression and a factor related to a certain expansion of the Bethe vector.
Such new factor and the presence of an inhomogeneous eigenvalue motivate
the nomenclature modified Slavnov and Gaudin-Korepin formulas for the models without $U(1)$ symmetry, to distinguish them from the diagonal or constrained cases known in the literature \cite{W02,KKMNST07I,KKMNST07II,Yang2}.
\\

\paragraph{\textbf{Basic formalism}}Let us recall the fundamental objects of quantum inverse scattering method for models on the segment \cite{Skly88}.
The main object is the R-matrix, solution of the Yang-Baxter equation,
\ben
R_{ab}(u-v)R_{ac}(u-w)R_{bc}(v-w)=R_{bc}(v-w)R_{ac}(u-w)R_{ab}(u-v),
\een
defined in the tensor product of complex vector spaces $V_a\otimes V_b\otimes V_c$ with $V_i=\mathbb C^2$.
In the case of the XXX spin chain, the associated $R$-matrix is called the rational $R$-matrix and it is given by,
\ben
R(u) = u +\, { P} \,,
\een 
where $P=\sigma^+\otimes\sigma^-+\sigma^-\otimes\sigma^++\frac{1}{2}(1+ \sigma^z\otimes\sigma^z)$ is the permutation operator\footnote{$
\sigma^{z}=\left(\begin{array}{cc}
       1 & 0\\
      0 & -1      \end{array}\right),\quad
\sigma^{+}=\left(\begin{array}{cc}
       0 & 1\\
      0 & 0      \end{array}\right),\quad
\sigma^{-}=\left(\begin{array}{cc}
       0 & 0\\
      1 & 0      \end{array}\right),\quad
      \sigma^{x}=\sigma^{+}+\sigma^{-}, \quad  \sigma^{y}=i(\sigma^{-}-\sigma^{+}) $.}. 
This $R$-matrix is $GL(2)$ invariant,
\ben\label{symQQ}
[R(u),\mathscr Q\otimes \mathscr Q]=0,
\een
 for any $\mathscr Q$ in $GL(2)$. 
For quantum integrable models on the segment additional objects must be considered, namely, the K-matrix, solution of the reflection equation,
\ben\label{re}
&&R_{ab}(u-v)K^-_{a}(u)R_{ab}(u+v)K^-_{b}(v)=K^-_{b}(v)R_{ab}(u+v)K^-_{a}(u)R_{ab}(u-v),
\een
and the dual K-matrix, solution of the dual reflection equation,
\ben\label{dre}
&&R_{ab}(-u+v)(K^+_{a}(u))^{t_a}R_{ab}(-u-v-2)(K^+_{b}(v))^{t_b}\nonumber\\&&\qquad=
(K^+_{b}(v))^{t_b}R_{ab}(-u-v-2)(K^+_{a}(u))^{t_a}R_{ab}(-u+v),
\een
where $t_i$ denotes the transposition in the $i$th space.
The most general scalar solutions of the equations (\ref{re}) and (\ref{dre}) are given by \cite{dVGR1},
\ben
K^{+}(u)=\left(\begin{array}{cc}
    q+u+1  & \xi^+(u+1) \\
    \xi^-(u+1) & q-u-1 
\end{array}\right)
, \quad
K^{-}(u)=\left(\begin{array}{cc}
    p+u & \eta^+ u \\
    \eta^- u & p-u 
\end{array}    \right),
\een
where $\{q,\xi^+,\xi^-,p,\eta^+,\eta^-\}\in\mathbb{C}^6$ are generic parameters.
Due to the $GL(2)$ invariance, we can choose $\eta^\pm=0$, which we will consider in the remaining of the text, without losing generality.
The R-matrix and the K-matrices allow us to construct the following transfer matrix ,
\ben
&&t(u) = \tr_{a}\big( K^{+}_{a}(u)K_{a}(u)\big) , \quad K_{a}(u)= T_{a}(u)\, K^{-}_{a}(u)\, \hat T_{a}(u)=
\left(\begin{array}{c c}
       \mA(u) & \mB(u)\\
       \mC(u) & \mD(u)+\frac{1}{2u+1}\mA(u)
      \end{array}
\right)_a\;,
\label{eqB}
\label{transfer}
\een
where the double-row monodromy matrix $K_{a}(u)$ is built by Sklyanin's dressing procedure for the $K^-$ matrix, and the bulk monodromy matrices
are given by,
\ben
&&T_{a}(u) = R_{a1}(u-\theta_{1}) \cdots R_{aN}(u-\theta_{N}),\qquad
\hat T_{a}(u) = R_{aN}(u+\theta_{N}) \cdots R_{a1}(u+\theta_{1}),
\een
with the free parameters $\{\theta_{1}, \ldots \theta_{N}\}$ called inhomogeneity parameters.
The operator entries of $K_a(u)$ act on the quantum space $\cH=\otimes_{i=1}^N \mathbb{C}^2$ and satisfy
commutation relations given in appendix \ref{com}. 
In terms of these operators, the transfer matrix reads,
\ben\label{trane}
t(u)=\alpha(u) \mA(u)+\delta(u)\mD(u)+\beta(u)\mB(u)+\gamma(u) \mC(u),
\een
where
\ben\label{coetransfer}
&&\alpha(u)=\phi(u)(q+u),\qquad \delta(u)=q-(u+1),\qquad
\beta(u)=\xi^{-}\,  (u+1),\qquad
\gamma(u)=\xi^{+}\,  (u+1),
\een
with
\ben
&&\phi(u)=\frac{2(u+1)}{2u+1}.
\een
Two transfer matrices at different values of their spectral parameter commute, \textit{i.e.}, $[t(u),t(v)]=0$.
Therefore, the expansion of the transfer matrix $t(u)$ with respect to the spectral parameter $u$ provides a set 
of commuting operators, among which the isotropic Heisenberg Hamiltonian, given by,
\ben\label{H}
&&H =   \frac{1}{q}\left(\sigma^{z}_{1} +\xi^+ \sigma^{+}_{1}+\xi^- \sigma^{-}_{1}\right)+\sum_{n=1}^{N-1}  
\Big( \sigma^x_{n} \otimes\sigma^x_{n+1} +\sigma^y_{n} \otimes\sigma^y_{n+1} +\sigma^z_{n} \otimes\sigma^z_{n+1}\Big )+ \frac{1}{p}\sigma^{z}_{N}
\label{Hamiltonian}.
\een
The main problem is thus to find the eigenvectors and eigenvalues of the transfer matrix. Finally, let us recall that, due to the diagonal form of the $K^-$ matrix,
there exist a highest weight vector and its dual, namely,
\ben\label{HWV}
|\Omega\rangle=\otimes_{i=1}^N\left(\begin{array}{c}1\\ 0\end{array} \right),\qquad
\langle\Omega|=\otimes_{i=1}^N\left(\begin{array}{cc}1 & 0\end{array} \right),
\een
such that,
\ben
&&\label{REP}\mA(u)|\Omega\rangle=\Lambda_1(u)|\Omega\rangle,\qquad
\mD(u)|\Omega\rangle=\Lambda_2(u)|\Omega\rangle,\qquad
\mC(u)|\Omega\rangle=0,\\
&&\label{REPdual}\langle\Omega|\mA(u)=\langle\Omega|\Lambda_1(u),\qquad\langle\Omega|\mD(u)=\langle\Omega|\Lambda_2(u),
\qquad\langle\Omega|\mB(u)=0,
\een
where
\ben\label{Lambda12}
&&\Lambda_1(u)=(u+p)\prod_{i=1}^N(u+1-\theta_i)(u+1+\theta_i),\qquad
\Lambda_2(u)=
\phi(-u-1)(p-u-1)\prod_{i=1}^N(u-\theta_i)(u+\theta_i).
\een
\vspace{0.1cm}
\paragraph{\textbf{Modified algebraic Bethe ansatz}} The Bethe vector for the present model
was firstly conjectured in \cite{BC13}, and we review its obtainment here. The key point
is to introduce a similarity transformation with a two-fold aim: to bring the original transfer matrix expression (\ref{trane})
to a modified diagonal form and to define a modified creation operator. Indeed, let us introduce,
\ben
&&\mathscr Q=\left(\begin{array}{c c}
 \xi^{+} & \rho\\
 -\rho & \xi^{-}
\end{array}
\right),\qquad\rho=1-\sqrt{1+\xi^{+}\xi^{-}}\,,
\een
which diagonalizes the $K^+$ matrix,
\ben\label{similarR}
&&\overbar K_a^{+}(u)=\mathscr Q_a^{-1}K_a^{+}(u)\mathscr Q_a=\left(\begin{array}{cc}
q+(1+u)(1-\rho)& 0 \\
0 & q-(1+u)(1-\rho)
\end{array}    \right)_a
\,,
\een
and, as a consequence, the transfer matrix (\ref{transfer}) can be
written as,
\ben\label{transferbar}
&&t(u)=\tr_a\left(\overbar{K}_a^{+}(u)\overbar{K}_a(u)\right),\quad
\overbar{K}_a(u)=\mathscr Q_a^{-1}K_a^{-}(u)\mathscr Q_a=
\left(\begin{array}{c c}
       \overline\mA(u) & \overline\mB(u)\\
       \overline\mC(u) & \overline\mD(u)+\frac{1}{2u+1}\overline\mA(u)
      \end{array}
\right)_a.
\een
The modified double-row monodromy operators are given explicitly by,
\ben\label{modOP}
&&\overline\mA(u)=\frac{1}{2(\rho-1)}\left(\left(\rho \phi(u)-2\right)\mA(u)+\rho\,\mD(u)-\xi^-\mB(u)-\xi^+\mC(u)\right),\nonumber\\
&&\overline\mD(u)=\frac{1}{2(\rho-1)}\left(\left(\rho \phi(-u-1)-2\right)\mD(u)+\rho\phi(u)\phi(-u-1)\mA(u)+\xi^-\phi(u)\mB(u)+\xi^+\phi(u)\mC(u)\right),\nonumber\\
&&\overline\mB(u)=\frac{1}{2(\rho-1)}\left(\xi^-\phi(-u-1)\mA(u)-\xi^-\,\mD(u)+\frac{\xi^-{}^2}{\rho}\mB(u)-\rho\,\mC(u)\right),\nonumber\\
&&\overline\mC(u)=\frac{1}{2(\rho-1)}\left(\xi^+\phi(-u-1)\mA(u)-\xi^+\,\mD(u)-\rho\,\mB(u)+\frac{\xi^+{}^2}{\rho}\mC(u)\right).
\een
Since $\overbar{K}_a^{+}(u)$ is a diagonal matrix, the transfer matrix acquires a modified diagonal form, namely,
\ben\label{transferdl}
t(u)=\bar\alpha(u) \overline\mA(u)+\bar\delta(u)\overline\mD(u),
\een
with
\ben
&&\bar\alpha(u)=\phi(u)(q+u(1-\rho)),\qquad
\bar\delta(u)=q-(1+u)(1-\rho).
\een
As a consequence of (\ref{symQQ}), the operators
$\{\overline\mA(u),\overline\mB(u),\overline\mC(u),\overline\mD(u)\}$
satisfy the same commutation relations as
$\{\mA(u),\mB(u),\mC(u),\mD(u)\}$, see
appendix \ref{com}. This means that the Bethe state
can be created by the action of the operator ($\overline\mC(u)$) $\overline\mB(u)$ on the (dual) highest vector  (\ref{HWV}).
Indeed, let us define\footnote{Hereafter, we will use the following compact notation. A set of $M$ variables $\{u_1,u_2,\dots,u_M\}$ is denoted by $\bar u $ with $\# \bar u =M$. If the $i$th element is removed, we indicate $\bar u_i=\bar u/u_i$ and,
if we also remove the element $u_j$, we denote $\bar u_{ij}=\bar u/\{u_i,u_j\}$. For products of functions (\textit{e.g.} $f(u,v)$) or of operators $\mathscr{O}$ (\textit{e.g.} $\overline\mB$), we use the convention
\ben
f(u,\bar u)=\prod_{i=1}^Mf(u,u_i), \quad f(\bar v,\bar u)=\prod_{i=1}^M\prod_{j=1}^Mf(v_j,u_i), \quad f(u_i,\bar u_i)=\prod_{j=1, j\neq i}^Mf(u_i,u_j),  \quad \mathscr{O}(\bar u)=\prod_{k=1}^M\mathscr{O}(u_k).\nonumber
\een},
\ben\label{rPsi}
&&|\Psi^M(\bar u)\rangle=
\overline\mB(u_1)\dots\overline\mB(u_M)|\Omega\rangle=
\overline\mB(\bar u)|\Omega\rangle
\een
and
\ben\label{lPsi}
&&\langle \Psi^M(\bar u)|=\langle\Omega|\overline\mC(u_1)\dots\overline\mC(u_M)
=\langle\Omega|\overline\mC(\bar u).
\een
The repeated use of the relations (\ref{ExchBB},\ref{ExchAB},\ref{ExchDB}) alows
us to obtain,
\ben\label{ADonSB}
&&\overline\mA(u)\overline\mB(\bar u)=
f(u,\bar u)\overline\mB(\bar u)\overline\mA(u)\nonumber\\
&&\qquad\qquad+
\sum_{i=1}^M g(u,u_i)f(u_i,\bar u_i)\overline\mB(\{u,\bar u_i\})\overline\mA(u_i)
+
w(u,u_i)h(u_i,\bar u_i)\overline\mB(\{u,\bar u_i\})\overline\mD(u_i),
\nonumber\\
&&\overline\mD(u)\overline\mB(\bar u)=
h(u,\bar u)\overline\mB(\bar u)\overline\mD(u)\nonumber\\
&&\qquad\qquad+\sum_{i=1}^M k(u,u_i)h(u_i,\bar u_i)\overline\mB(\{u,\bar u_i\})\overline\mD(u_i)
+ n(u,u_i)f(u_i,\bar u_i)\overline\mB(\{u,\bar u_i\})\overline\mA(u_i),
\een
while the use of (\ref{ExchCC},\ref{ExchCA},\ref{ExchCD}) gives us,
\ben\label{SConAD}
&&\overline\mC(\bar u)\overline\mA(u)=
f(u,\bar u)\overline\mA(u)\overline\mC(\bar u)\nonumber\\
&&\qquad\qquad+
\sum_{i=1}^M g(u,u_i)f(u_i,\bar u_i)\overline\mA(u_i)\overline\mC(\{u,\bar u_i\})
+
w(u,u_i)h(u_i,\bar u_i)\overline\mD(u_i)\overline\mC(\{u,\bar u_i\}),
\nonumber\\
&&\overline\mC(\bar u)\overline\mD(u)=
h(u,\bar u)\overline\mD(u)\overline\mC(\bar u)\nonumber\\
&&\qquad\qquad+
\sum_{i=1}^M k(u,u_i)h(u_i,\bar u_i)\overline\mD(u_i)\overline\mC(\{u,\bar u_i\})
+
n(u,u_i)f(u_i,\bar u_i)\overline\mA(u_i)\overline\mC(\{u,\bar u_i\}).
\een
In order to evaluate the action of the transfer matrix on (\ref{rPsi}) and (\ref{lPsi}), in addition to (\ref{ADonSB}) and (\ref{SConAD}), we also need
to compute the action of the operators $\overline\mA(u)$ and $\overline\mD(u)$ on (\ref{HWV}). It is a modified action, namely,
\ben\label{modREP}
&&\overline\mA(u)|\Omega\rangle=\Lambda_1(u)|\Omega\rangle-\frac{\rho}{\xi^-}\overline\mB(u)|\Omega\rangle,\qquad
\overline\mD(u)|\Omega\rangle=\Lambda_2(u)|\Omega\rangle+\frac{\rho}{\xi^-}\phi(u)\overline\mB(u)|\Omega\rangle,\nonumber\\
&&\overline\mC(u)|\Omega\rangle=
\frac{\rho}{\xi^-}\left(\phi(-u-1)\Lambda_1(u)-\Lambda_2(u)\right)|\Omega\rangle-\left(\frac{\rho}{\xi^-}\right)^2\overline\mB(u)|\Omega\rangle,\\
&&\label{modREPdual}\langle\Omega|\overline\mA(u)=\Lambda_1(u)\langle\Omega|-\frac{\rho}{\xi^+}\langle\Omega|\overline\mC(u),\qquad
\langle\Omega|\overline\mD(u)=\Lambda_2(u)\langle\Omega|+\frac{\rho}{\xi^+}\phi(u)\langle\Omega|\overline\mC(u),\nonumber\\
&&\langle\Omega|\overline\mB(u)=
\frac{\rho}{\xi^+}\left(\phi(-u-1)\Lambda_1(u)-\Lambda_2(u)\right)\langle\Omega|-\left(\frac{\rho}{\xi^+}\right)^2\langle\Omega|\overline\mC(u).
\een
Using the multiple actions (\ref{ADonSB},\ref{SConAD}) together with
(\ref{modREP},\ref{modREPdual}), we obtain,
\ben
&&t(u)|\Psi^M(\bar u)\rangle=
\Lambda_{d}^M(u,\bar u)|\Psi^M(\bar u)\rangle
+\sum_{i=1}^MF(u,u_i)E_{d}^M(u_i,\bar u_i)|\Psi^M(\{u,\bar u_i\})\rangle
\nonumber\\
&&\qquad+\frac{\rho(\rho-1)}{\xi^-}2(u+1)\,\overline\mB(u)|\Psi^M(\bar u)\rangle,\label{OFFpartial}\\
&&\label{OFFpartialleft}\langle \Psi^M(\bar u)|t(u)=
\langle \Psi^M(\bar u)|\Lambda_{d}^M(u,\bar u)
+\sum_{i=1}^MF(u,u_i)E_{d}^M(u_i,\bar u_i)\langle \Psi^M(\{u,\bar u_i\})|
\nonumber\\\
&&\qquad+\frac{\rho(\rho-1)}{\xi^+}2(u+1)\langle \Psi^M(\bar u)|\overline\mC(u),
\een
where
\ben\label{Lamd}
&&\Lambda_{d}^{M}(u,\bar u)=\bar\alpha(u)\Lambda_1(u)f(u,\bar u)+
\bar\delta(u)\Lambda_2(u)h(u,\bar u),\nonumber\\
&&
E_{d}^M(u_i,\bar u_i)=-\phi(-u_i-1)\bar\alpha(u_i)\Lambda_1(u_i)f(u_i,\bar u_i)+
\phi(u_i)\bar\delta(u_i)\Lambda_2(u_i)h(u_i,\bar u_i),
\een
with
\ben
F(u,v)=-\frac{\phi(u)(2u+1)}{\phi(v)Q(u,v)},\qquad Q(u,v)=(u-v)(u+v+1).
\een
The new terms in (\ref{OFFpartial},\ref{OFFpartialleft}), \textit{i.e.}, the proportional terms to $\overline\mB(u)|\Psi^M(\bar u)\rangle$ and $\langle \Psi^M(\bar u)|\overline\mC(u)$, are characteristic of MABA approach. If the number
of creation operators equals the length of the chain, \textit{i.e.}, $M=N$, these terms are given by,
\ben\label{OFF}
&&\frac{\rho(\rho-1)}{\xi^-}2(u+1)\,\overline\mB(u)|\Psi^N(\bar u)\rangle=
\Lambda_{g}^N(u,\bar u)|\Psi^N(\bar u)\rangle+
\sum_{i=1}^NF(u,u_i)E_{g}^N(u_i,\bar u_i)|\Psi^N(\{u,\bar u_i\})\rangle,\\
&&\label{OFFleft}\frac{\rho(\rho-1)}{\xi^+}2(u+1)\,\langle \Psi^N(\bar u)|\overline\mC(u)=
\langle \Psi^N(\bar u)|\Lambda_{g}^N(u,\bar u)+
\sum_{i=1}^NF(u,u_i)E_{g}^N(u_i,\bar u_i)\langle \Psi^N(\{u,\bar u_i\})|,
\een
with
\ben\label{Lamg}
&&\Lambda_{g}^N(u,\bar u)=
\rho\,\tilde\phi(u)
\frac{\Lambda_1(u)\Lambda_2(u)}{Q(u,\bar u)},\quad
E_{g}^N(u_i,\bar u_i)=
\rho\,\frac{\tilde\phi(u_i)}{2u_i+1}
\frac{\Lambda_1(u_i)\Lambda_2(u_i)}{Q(u_i,\bar u_i)},\quad \tilde\phi(u)=\frac{(u+1)(2u+1)}{(p+u)(p-u-1)},
\een
and $\# \bar u=N$.
The equations (\ref{OFF},\ref{OFFleft}) are the central relations in the MABA. Their proof follows from the rational limit of the proof given in \cite{ABGP15} for the XXZ case.
We arrive, combining (\ref{OFFpartial}) and (\ref{OFF}), to the final off-shell equation
satisfied by the left and right Bethe vectors,
\ben
&&t(u)|\Psi^N(\bar u)\rangle=
\Lambda^N(u,\bar u)|\Psi^N(\bar u)\rangle
+\sum_{i=1}^NF(u,u_i)E^N(u_i,\bar u_i)|\Psi^N(\{u,\bar u_i\})\rangle,\label{OFFPsi}\\
&&\label{OFFPsileft}\langle \Psi^N(\bar u)|t(u)=
\langle \Psi^N(\bar u)|\Lambda^N(u,\bar u)
+\sum_{i=1}^NF(u,u_i)E^N(u_i,\bar u_i)
\langle \Psi^N(\{u,\bar u_i\})|,
\een
where the eigenvalue and the corresponding Bethe equations are given by,
\ben\label{Lam}
&&\Lambda^N(u,\bar u)=\Lambda_{d}^N(u,\bar u)+\Lambda_g^N(u,\bar u),\qquad
E^N(u_i,\bar u_i)=E_{d}^N(u_i,\bar u_i)+E_{g}^N(u_i,\bar u_i)
\een
with $\# u=N$. We recall here that the eigenvalue expression (\ref{Lam}) was firstly obtained in \cite{ODBAXXX}, and it was further developed in \cite{Nepo13}.
\vspace{0.1cm}
\paragraph{\textbf{Modified Slavnov and Gaudin-Korepin formulas}} We are now in position to investigate
the scalar product between the Bethe vectors and its dual, namely,
\ben
S^N(\bar u|\bar v)=\langle \Psi^N(\bar u)|\Psi^N(\bar v)\rangle.
\een
Considering $\langle \Psi^N(\bar u)|$ on-shell, \textit{i.e.},
$E^N(u_i,\bar u_i)=0$ for $i=1,\dots N$,
we conjecture
the following modified Slavnov formula,
\ben\label{modSLAV}
\hat S^N(\bar u,\bar v)=\left(\frac{\rho-2}{2(\rho-1)^2}\right)^N W_0^N(\bar u) \frac{{\textrm{Det}_N\left(\frac{\partial}{\partial u_i}\Lambda^N(v_j,\bar u)\right)}}{\textrm{Det}_N\left(V(v_i,u_j)\right)},
\een
where
\ben
W_0^N(\bar u)=\left(\frac{2(\rho-1)}{\xi^-}\right)^N\langle 0|\overline\mB(\bar u)|0\rangle,\qquad
V(v_i,u_j)=\frac{(2u_j+1)2(v_i+1)}{Q(v_i,u_j)}\,,
\een
with $\# \bar u=N$. The factor $W_0^N(\bar u)$ is related to the expansion of the Bethe vector in the basis of the original
operators $\{\mA(u),\mB(u),\mC(u),\mD(u)\}$ and it is given in appendix \ref{appB}.
If, on the other hand, $|\Psi^N(\bar v)\rangle$ is on-shell, namely $E^N(v_i,\bar v_i)=0$ for $i=1,\dots N$, the respective modified formula is obtained by interchanging $u_i\leftrightarrow v_i$ in
the right-hand side of (\ref{modSLAV}),
\ben\label{modSLAVv}
\tilde S(\bar u,\bar v)=\left(\frac{\rho-2}{2(\rho-1)^2}\right)^N W_0^N(\bar v) \frac{{\textrm{Det}_N\left(\frac{\partial}{\partial v_i}\Lambda^N(u_j,\bar v)\right)}}{\textrm{Det}_N\left(V(u_i,v_j)\right)}.
\een
The modified Gaudin-Korepin formula can be obtained by taking the limit $\bar v=\bar u$ in (\ref{modSLAV}). In order to do that
we firstly rewrite the Cauchy-like determinant as,
\ben
&&\textrm{Det}_N\left(V(v_i,u_j)\right)=\frac{\prod_{i=1}^N(2u_i+1)2(v_i+1)\prod_{i<j}^NQ(u_j,u_i)Q(v_i,v_j)}
{\prod_{i,j=1}^NQ(v_j,u_i)},
\een
and, next, place the term $\prod_{i,j=1}^NQ(v_j,u_i)$ along with the Jacobian. We can then perform the limit,
\ben
\lim_{v_j\to u_j} \left(\prod_{k=1}^NQ(v_j,u_k)\frac{\partial}{\partial u_i}\Lambda^N(v_j,\bar u)\right),
\een
using l'Hospital's rule. As a result we obtain the following
expression for the square of the norm,
\ben
{\mathscr N}^N(\bar u)=\left(\frac{\rho-2}{2(\rho-1)^2}\right)^N W_0(\bar u)
\frac{\textrm{det}_N\left(G_{ij}\right)}{\prod_{i=1}^N2(u_i+1)\prod_{i<j}^NQ(u_j,u_i)Q(u_i,u_j)},
\een
where the matrix elements $G_{ij}$ for $i,j=1,\dots N$ are given by,
\ben\label{modGaud}
&&G_{ii}=Q(u_i,\bar u_i)\frac{\partial E^N(u_i,\bar u_i)}{\partial u_i},\nonumber\\
&&G_{ij}=(2u_j+1)\left(\phi(-u_j-1)\bar\alpha(u_j)\Lambda_1(u_j)Q(-u_j,\bar u_{ij})
-\phi(u_j)\bar\delta(u_j)\Lambda_2(u_j)Q(u_j+1,\bar u_{ij})\right),\nonumber\\
&&\qquad\qquad\qquad\qquad\qquad\qquad\qquad\qquad\qquad\qquad\qquad\qquad\qquad\qquad\qquad\qquad\qquad\qquad\textrm{for}\quad i\neq j.
\een
Evaluating the derivative in $G_{ii}$ and using the Bethe equations, we obatin the following explicit expression,
\ben
&&G_{ii}=
-\phi(-u_i-1)\bar\alpha(u_i)\Lambda_1(u_i)Q(-u_i,\bar u_i)\left((2u_i-1)\sum_{k\neq i}^N\frac{1}{Q(-u_i,u_k)}+\left(\frac{1}{u_i}-\frac{\partial_{u_i}\tilde\phi(u_i)}{\tilde\phi(u_i)}+\frac{\partial_{u_i}\bar\alpha(u_i)}{\bar\alpha(u_i)}\right)\right)\nonumber\\&&
\qquad+\phi(u_i)\bar\delta(u_i)\Lambda_2(u_i)Q(u_i+1,\bar u_i)\left((2u_i+3)\sum_{k\neq i}^N\frac{1}{Q(u_i+1,u_k)}+\left(\frac{1}{u_i+1}-\frac{\partial_{u_i}\tilde\phi(u_i)}{\tilde\phi(u_i)}+\frac{\partial_{u_i}\bar\delta(u_i)}{\bar\delta(u_i)}\right)\right)\nonumber\\
&&\qquad-
\frac{\partial \Lambda_1(u_i)}{\partial u_i}\left(\phi(-u_i-1)\bar\alpha(u_i)Q(-u_i,\bar u_i)-\frac{\rho\tilde\phi(u_i)\Lambda_2(u_i)}{2u_i+1}\right)\nonumber\\
&&\qquad+\frac{\partial \Lambda_2(u_i)}{\partial u_i}\left(\phi(u_i)\bar\delta(u_i)Q(u_i+1,\bar u_i)+\frac{\rho\tilde\phi(u_i)\Lambda_1(u_i)}{2u_i+1}\right).
\een

\paragraph{\textbf{Construction of the conjecture}}

To obtain the conjecture of the modified Slavnov formula in the previous paragraph, we have used the same procedure from the twisted XXX chain case considered in \cite{BelPimtwist}. The construction of, for instance (\ref{modSLAVv}), is based on the following line of reasoning:
\begin{enumerate}[I.]
\item The modified formula (\ref{modSLAVv}) must be reduced to the formula of the case with diagonal boundaries ($\rho=0$), which is known in the literature \cite{W02,KKMNST07I,KKMNST07II}. In our notation, it reads,
\ben\label{modSLAVd}
&&\tilde S_d^M(\bar u,\bar v)=\Lambda_2(\bar v)\prod_{i=1}^M\frac{2 v_i+1}{v_i+q} \prod_{j<i}\frac{v_i+v_j+2}{v_i+v_j}  \frac{{\textrm{Det}_M\left(\frac{\partial}{\partial v_i}\Lambda_d^M(u_j,\bar v)\right)}}{\textrm{Det}_M\left(V(u_i,v_j)\right)},
\een
where $v_i$ for $i=1,\dots,M$ satisfies $E_{d}^M(v_i,\bar v_i)=0$. From this, we expect that the modified formula contains the Jacobian of the inhomogeneous eigenvalue $\Lambda^N(u,\bar v)$ given by (\ref{Lam}) and an additional factor to replace $\Lambda_2(\bar v)$, namely,
\ben\label{ansatz}
\tilde S^N(\bar u,\bar v)=\zeta^N\, Z^N(\bar u,\bar v)\frac{{\textrm{Det}_N\left(\frac{\partial}{\partial v_i}\Lambda^N(u_j,\bar v)\right)}}{\textrm{Det}_N\left(V(u_i,v_j)\right)},
\een
where the constant $\zeta$ and the function $Z^N(\bar u,\bar v)$ are to be fixed.
\item We recall that the factor $\Lambda_2(\bar v)$ in (\ref{modSLAVd}) comes from using, in some expression for the off-shell scalar product, the Bethe
equations without solving it explicitly, \textit{i.e.}, by writing $\Lambda_1(v_i)\sim\Lambda_2(v_i)$ from $E_{d}^M(v_i,\bar v_i)=0$  in (\ref{Lamd}).
For the general model, since the Bethe equations (\ref{Lam}) are quadratic in $\Lambda_{1,2}(v_i)$, the only way to use the Bethe equations is by expressing the term $\Lambda_{1}(v_i)\Lambda_{2}(v_i)$ through (\ref{Lam})
as follows,
\ben\label{linear}
&&\Lambda_{1}(v_i)\Lambda_{2}(v_i)=\frac{(2v_i+1)Q(v_i,\bar v_i)}{\rho\tilde\phi(v_i)}\left(\phi(-v_i-1)\bar\alpha(v_i)\Lambda_1(v_i)f(v_i,\bar v_i)-
\phi(v_i)\bar\delta(v_i)\Lambda_2(v_i)h(v_i,\bar v_i)\right),
\een
for $i=1,\dots,N$.
\item The function $Z^N(\bar u,\bar v)$ can be fixed from a recursion relation on the scalar product up to the constant $\zeta$.  
\item The constant $\zeta$ can be fixed from the case $N=1$ which we consider explicitly. 
\end{enumerate}
Considering the above points, let us fix the function $Z^N(\bar u,\bar v)$ through the action of the $\overline\mC(u)$ over the Bethe vector (\ref{rPsi}),
namely,
\ben\label{CPSI}
\overline\mC(u)|\Psi^N(\bar v)\rangle&=&
-\Big(\frac{\rho}{\xi^-}\Big)^2\,\overline{\mB}(u)|\Psi^N(\bar v)\rangle
+\frac{\rho}{\xi^-}
\left(\phi(-u-1)\Lambda_1(u)f(u,\bar v)
-\Lambda_2(u)h(u,\bar v)\right)|\Psi^N(\bar v)\rangle
\nonumber\\\nonumber&&-\frac{\rho}{\xi^-}\,
\sum_{i=1}^N
w(u,v_i)\Big(2v_i\Lambda_1(v_i)g(u,v_i)f(v_i,\bar v_i)\\
&&\qquad +(1+2v_i)\Lambda_2(v_i)k(v_i,u)h(v_i,\bar v_i)
\Big)
|\Psi^N(\{u,\bar v_i\})\rangle\nonumber\\&&\qquad\qquad+
\sum_{i=1}^NH_i(u,\bar v)|\Psi^{N-1}(\bar v_i)\rangle
+\sum_{i<j}^NH_{ij}(u,\bar v)|\Psi^{N-1}(\{u,\bar v_{ij}\})\rangle,
\een
where the auxiliary functions $H_i(u,\bar v)$ and $H_{ij}(u,\bar v)$ are given in appendix \ref{com}. Taking into account the off-shell
action (\ref{OFF}), the equation (\ref{CPSI}) leads to,
\ben\label{asymptotic}
&&S^N(\bar u|\bar v)\sim\prod_{i=1}^N\frac{\Lambda_g(u_i,\bar v)}{2(u_i+1)}\langle 0|\Psi^N(\bar v)\rangle+\cdots
\een
where we identify a leading term with $N$ quadratic terms: $\Lambda_1(u_i)\Lambda_2(u_i)$ with $i=1,\dots N$. This term
is the only invariant one, of order $3N$ in the $\Lambda_{1,2}(v_i)$,  under the prescription (\ref{linear}). This suggests to us to identify the unknown function in (\ref{ansatz}) as,
\ben
Z^N(\bar u,\bar v)=W_{0}^N(\bar v).
\een
Indeed, the leading term of the Jacobian in $\Lambda_1(v_i)\Lambda_2(v	_i)$ is given by,
\ben
{\textrm{Det}_N\left(\frac{\partial}{\partial v_i}\Lambda_g^N(u_j,\bar v)\right)}=\prod_{i=1}^N \frac{\Lambda_g(u_i,\bar v)}{2(u_i+1)}{\textrm{Det}_N\left(V(u_i,v_j)\right)},
\een
which makes the asymptotic behavior of (\ref{ansatz}) consistent with (\ref{asymptotic}). Let us also note that in the diagonal boundaries limit, $\rho=0$, we have
\ben
W_{0}^N(\bar v)=\prod_{i=1}^N\frac{-2 v_i-1}{v_i+q} \Lambda_2(v_i) \prod_{i<j}\frac{v_i+v_j+2}{v_i+v_j},
\een
and the usual formula (\ref{modSLAVd}) is recovered, up to a constant. The remaining task is to fix the constant $\zeta$ in (\ref{ansatz}).
In order to do that, we need to find a good parametrization for off-shell scalar product such that the prescription (\ref{linear}) can be used.
In the simplest case, $N=1$, it turns out that the off-shell scalar product can be written as,
\ben\label{goodS1}
S^1(u_1|v_1)=\frac{\rho-2}{2(\rho-1)^2}\left((\rho-1)S_d^1(u_1|v_1)+\left(\frac{\Lambda_g^1(u_1,v_1)W_0^1(v_1)}{2(u_1+1)}+\frac{\Lambda_g^1(v_1,u_1)W_0^1(u_1)}{2(v_1+1)}\right)\right)
\een
where
\ben
&&S_d^1(u_1|v_1)=(s(u_1,v_1)+x(u_1,v_1)) \Lambda_1(u_1) \Lambda_1(v_1)
+y(u_1,v_1) \Lambda_2(u_1) \Lambda_1(v_1)
+r(u_1,v_1)\Lambda_1(u_1)\Lambda_2(v_1)
\nonumber\\
&&\qquad\qquad+q(u_1,v_1) \Lambda_1(v_1)\Lambda_2(u_1)+w(u_1,v_1) \Lambda_2(u_1) \Lambda_2(v_1),
\een
is the diagonal contribution to the scalar product. In the form (\ref{goodS1}), the use of (\ref{linear}) leads us directly to,
\ben\label{SLAVfromgoodS1}
\tilde S^1(u_1|v_1)=\frac{\rho-2}{2(\rho-1)^2}W_0^1(v_1)\frac{\frac{\partial}{\partial v_1}\Lambda^1(u_1,v_1)}{V(u_1,v_1)},
\een
fixing thus the desired constant. To find a convenient off-shell representation for the scalar product for general $N$ that allows us to use (\ref{ansatz})
remains an open problem.  For example, the off-shell scalar product obtained from the projection of the Bethe vector on the diagonal operator basis, see appendix \ref{appB},  gives us,
\ben\label{s1}
S^1(u_1|v_1)=\left(\frac{\rho-2}{2(\rho-1)}\right)^2S_d^1(u_1|v_1)+\frac{\rho(\rho-2)}{(2(\rho-1))^2}W_0^1(u_1)W_0^1(v_1).
\een
Here the use of the prescription (\ref{linear}) for the quadratic term cannot be applied. 
Alternatively, from the use of the action (\ref{CPSI}), we obtain,
\ben\label{s2}
&&S^1(u_1|v_1)=S_d^1(u_1|v_1)-\frac{\rho}{2(\rho-1)^2}\left(\frac{\Lambda_g^1(u_1,v_1)W_0^1(v_1)}{2(u_1+1)}+\frac{\Lambda_g^1(v_1,u_1)W_0^1(u_1)}{2(v_1+1)}\right)\nonumber\\
&&+\frac{\rho}{2(\rho-1)}\Bigg(\phi(-v_1-1)\Lambda_1(v_1)\left(\frac{u_1+v_1-1}{u_1+v_1+1}\phi(-u_1-1)\Lambda_1(u_1)-\frac{u_1-v_1+2}{u_1-v_1}\Lambda_2(u_1)\right)\nonumber\\&&-\Lambda_2(v_1)\left(\frac{u_1-v_1-2}{u_1-v_1}\phi(-u_1-1)\Lambda_1(u_1)-\frac{u_1+v_1+3}{u_1+v_1+1}\Lambda_2(u_1)\right)\Bigg),
\een
where we observe the presence of quadratic terms that can be simplified by (\ref{linear}). However, it does not lead directly to the modified Slavnov formula due to
the presence of additional contributions. We remark that all the representations (\ref{goodS1},\ref{s1},\ref{s2}) are the same, as expected, when
we use the explicit form of $\Lambda_{1,2}(u)$.

By any means, the $N=1$ case is already enough to obtain the constant and thus to propose the general formulas.
We have numerically verified that both (\ref{modSLAV},\ref{modSLAVv}) and (\ref{modGaud}) are valid for $N=2,3$.
\vspace{0.1cm}

\paragraph{\textbf{Discussion}} We have obtained, in the framework of the modified algebraic Bethe ansatz, a compact expression for the scalar product
between an on-shell Bethe vector and its off-shell dual for the isotropic spin-$1/2$ Heisenberg chain on the segment with general integrable boundaries.
It is a modified Slavnov formula, in the sense that new factors appear, when compared with the diagonal boundary or periodic cases.
Many interesting questions remain to be investigated. The first one is to obtain a proof
of our conjecture. Next, we should consider the evaluation of form factors and correlations functions of the model. The XXZ case can be also considered, and the result has the same form of (\ref{modSLAV},\ref{modSLAVv}), with $\Lambda^N(u,\bar u)$ and $\textrm{Det}_N(V(u,v))$ replaced by the corresponding $q-$deformed versions. The factor $W_0^N(\bar u)$ will be given by (\ref{defW}) with $W^{i}_{i-1}(u_{1}| u_{2}, \dots ,u_{i})=\Lambda_{ps}^{i-1}(u_1,\{u_{2}, \dots ,u_{i}\})$, using the notaion of \cite{BelPim15}. It remains to find the overall constant and to prove the conjecture.

\vspace{0.2cm}

\paragraph{\textbf{Acknowledgements}} We would like to thank R. Nepomechie for discussions. R.A.P would like to thanks the hospitality of the Laboratoire de Physique Th\'{e}orique et Mod\'{e}lisation at
the Universit\'e de Cergy-Pontoise where a part of this work was done. S.B. is supported by the Universit\'e de Cergy-Pontoise post doctoral fellowship and by a public grant as part of the
Investissement d'avenir project, reference ANR-11-LABX-0056-LMH,
LabEx LMH.  R.A.P. is supported by Sao Paulo Research Foundation (FAPESP), grants \# 2014/00453-8 and \# 2014/20364-0.

\appendix

\section{Commutation relations and functions \label{com}}
Using reflection equation \eqref{re}, we can obtain the exchange relations between the operators $\{\mA,
\mB, \mC, \mD\}$. A list of relevant ones for this work is,
\ben
\mB(u)\mB(v)&=&\mB(v)\mB(u)\,, \label{ExchBB}
\een
\ben
\mC(u)\mC(v)&=&\mC(v)\mC(u)\,,\label{ExchCC}\\
\mA(u)\mB(v)&=& f(u,v)\mB(v)\mA(u) + g(u,v)\mB(u)\mA(v) + w(u,v)\mB(u)\mD(v)\;, \label{ExchAB}\\
\mC(v)\mA(u)&=& f(u,v)\mA(u) \mC(v)+ g(u,v)\mA(v) \mC(u)+ w(u,v)\mD(v)\mC(u)\;,\label{ExchCA} \\
\mD(u)\mB(v)&=& h(u,v)\mB(v) \mD(u) +k(u,v)\mB(u)\mD(v)+ n(u,v)\mB(u)\mA(v)\;,\label{ExchDB}\\
\mC(v)\mD(u)&=& h(u,v) \mD(u)\mC(v) +k(u,v)\mD(v)\mC(u)+ n(u,v)\mA(v)\mC(u)\;,\label{ExchCD}\\
\mC(u)\mB(v)&=&\mB(v)\mC(u)+s(u,v) \mA(u) \mA(v)+x(u,v) \mA(v) \mA(u)+y(u,v) \mD(u) \mA(v)\label{ExchCB}\nonumber\\
&&+r(u,v)\mA(u)\mD(v)+q(u,v) \mA(v)\mD(u)+w(u,v) \mD(u) \mD(v)\;,
\een
where
\ben
&&f(u,v)= \frac{(u-v-1)(u+v)}{(u-v)(u+v+1)},\qquad g(u,v)= \frac{2v}{(2v+1)(u-v)},\qquad w(u,v)= \frac{-1}{(u+v+1)},\nonumber\\
&& h(u,v)= \frac{(u-v+1)(u+v+2)}{(u-v)(u+v+1)},\quad
k(u,v)= \frac{-2(u+1)}{(u-v)(2u+1)},\quad n(u,v)= \frac{4v(u+1)}{(u+v+1)(2v+1)(2u+1)},
\een
and
\ben
&& x(u,v)=\frac{2\,  u\,(u-v+1)}{(2\,u+1)(u+v+1)(u-v)},\qquad s(u,v)=-\frac{2 \,u}{(2\,u+1)(2\,v+1)(u-v)},\nonumber\\
&& q(u,v)=\frac{(u+v)}{(u+v+1)(u-v)},\qquad r(u,v)=-\frac{2 \,u}{(2\,u+1)(u-v)}, \qquad
y(u,v)=-\frac{1}{(u+v+1)(2v+1)}.
\een
Along the main text we use frequently the functions,
\ben
\phi(u)=\frac{2(u+1)}{2u+1},\qquad 
F(u,v)=-\frac{\phi(u)(2u+1)}{\phi(v)Q(u,v)},\qquad Q(u,v)=(u-v)(u+v+1).
\een
In addition, we have the following auxiliary functions entering equation (\ref{CPSI}),
\ben\label{defHk}
&&H_{k}(u,\bar u)=\Lambda_{1}(u)\Big(\Lambda_{1}(u_{k})\big(
s(u,u_{k})+x(u,u_{k})\big) f(u,\bar u_k)f(u_k,\bar u_k)+
\Lambda_{2}(u_{k})r(u,u_{k})f(u,\bar u_k)h(u_k,\bar u_k) \Big) \nonumber   \\
&&\qquad +\Lambda_{2}(u)\Big(\Lambda_{1}(u_{k})\big( q(u,u_{k})+y(u,u_{k})
\big) h(u,\bar u_k)f(u_k,\bar u_k)
+\Lambda_{2}(u_{k})w(u,u_{k})h(u,\bar u_k)h(u_k,\bar u_k)\Big)
\een

\ben\label{defHlll}
&& H_{kl}(u,\bar u)=
\Lambda_{1}(u_{k})\Big(\Lambda_{1}(u_{l})\alpha _{11}(u,u_{k},u_{l})f(u_k,\bar u_{kl})f(u_l,\bar u_{kl})+
\Lambda_{2}(u_{l})\alpha _{12}(u,u_{k},u_{l})f(u_k,\bar u_{kl})h(u_l,\bar u_{kl})\Big) \nonumber
\\
&&\qquad +
\Lambda_{2}(u_{k })\Big(\Lambda_{1}(u_{l})\alpha _{21}(u,u_{k},u_{l})
h(u_k,\bar u_{kl})f(u_l,\bar u_{kl})
+
\Lambda_{2}(u_{l})\alpha _{22}(u,u_{l},u_{k})
h(u_k,\bar u_{kl})h(u_l,\bar u_{kl})\Big)
\een
with
\ben
&&\alpha _{11}(u,u_{k},u_{l})\nonumber\\&&\qquad = g(u,u_l) (s(u,u_k)
   f(u_k,u_l)+f(u_k,u)
   x(u,u_k))+n(u,u_l) (y(u,u_k)
   f(u_k,u_l)+f(u_k,u)
   q(u,u_k))\nonumber\\
&&
   \qquad +g(u,u_k) (s(u,u_k)
   g(u_k,u_l)+r(u,u_k)
   n(u_k,u_l))+n(u,u_k) (y(u,u_k)
   g(u_k,u_l)+w(u,u_k)
   n(u_k,u_l)),
\een
\ben
&&\alpha _{12}(u,u_{k},u_{l})\nonumber\\&&\qquad =k(u,u_{l}) (f(u_{k},u)
   q(u,u_{k})+f(u_{k},u_{l})
   y(u,u_{k}))+w(u,u_{l})
   (f(u_{k},u_{l})
   s(u,u_{k})+f(u_{k},u)
   x(u,u_{k}))\nonumber\\&&\qquad +g(u,u_{k})
   (k(u_{k},u_{l})
   r(u,u_{k})+s(u,u_{k})
   w(u_{k},u_{l}))+n(u,u_{k})
   (k(u_{k},u_{l})
   w(u,u_{k})+y(u,u_{k})
   w(u_{k},u_{l})),
\een
\ben
&&\alpha _{21}(u,u_{k},u_{l})\nonumber\\&&\qquad =r(u,u_{k}) (g(u,u_{l})
   h(u_{k},u_{l})+n(u_{k},u_{l}) w(u,u_{k}))+g(u_{k},u_{l})
   (k(u,u_{k})
   y(u,u_{k})+s(u,u_{k})
   w(u,u_{k}))\nonumber\\&&\qquad +w(u,u_{k})
   (h(u_{k},u_{l})
   n(u,u_{l})+k(u,u_{k})
   n(u_{k},u_{l})),
\een
\ben
&&\alpha _{22}(u,u_{k},u_{l})\nonumber\\&&\qquad=r(u,u_{k}) (h(u_{k},u_{l})
   w(u,u_{l})+k(u_{k},u_{l})
   w(u,u_{k}))+w(u,u_{k})
   (h(u_{k},u_{l})
   k(u,u_{l})+k(u,u_{k})
   k(u_{k},u_{l}))\nonumber\\&&\qquad +w(u_{k},u_{l}) (k(u,u_{k})
   y(u,u_{k})+s(u,u_{k})
   w(u,u_{k})).
\een

\section{Projection of the Bethe vector \label{appB} }
Using the commutation relations given in appendix \ref{com} and the representation
theory (\ref{REP}), the Bethe vector (\ref{rPsi}) can be expanded in terms of the operator $\mB(u)$ in the following way,
\ben\label{expPSI}
\overline\mB(\bar u)|0\rangle= \left(\frac{(\rho-2)\xi^-}{2(\rho-1)\xi^+}\right)^N
\sum_{i=0}^{N} \sum_{\bar u \to \{\bar u_{\so},\bar u_{\st}\}}\left(\frac{\rho}{\xi^-}\right)^{N-i} \, W^{N}_{i}(\bar u_{\so}|\bar u_{\st})\mB(\bar u_{\st})|0\rangle
\een
with  $\#\bar u_{\st}=i$, $\#\bar u_{\so}=N-i$ a partition of the set $\bar u$. The sum is over all ordered partitions, denoted $\bar u \to \{\bar u_{\so},\bar u_{\st}\}$. The coefficient is given by   
\ben\label{defW}
&& W^{N}_{i}(u_{1}, \dots , u_{N-i}|u_{N-i+1},\dots , u_{N})=\mbox{Sym}^{N-i}_{u_{1}, \dots , u_{N-i}}\Big(\prod_{j=1}^{N-i}W^{N+1-j}_{N-j}(u_{j}| u_{j+1}, \dots ,u_{N})\Big)
\een
where
\ben
W^{i}_{i-1}(u_{1}| u_{2}, \dots ,u_{i})=\phi(-u_1-1)\Lambda_1(u_1)f(u_1,\bar u_1)-\Lambda_2(u_1)h(u_1,\bar u_1),
\een
and
\ben
\mbox{Sym}^{M}_{\bar u}\Big(F(\bar u)\Big)=\frac{1}{M!}\sum_{\sigma \in S_M}F(\bar u^\sigma)
\een
with $\bar u^\sigma=\{u_{\sigma(1)},\dots,u_{\sigma(M)}\}$ an element of the permutation group $S_M$. In the same way,
the dual Bethe vector (\ref{lPsi}) in terms of the operator $\mC(u)$ is given by,
\ben\label{expPSIleft}
\langle 0|\overline\mC(\bar u)=\left(\frac{(\rho-2)\xi^+}{2(\rho-1)\xi^-}\right)^N
\sum_{i=0}^{N} \sum_{\bar u \to \{\bar u_{\so},\bar u_{\st}\}}
\left(\frac{\rho}{\xi^+}\right)^{N-i} W^{N}_{i}(\bar u_{\so}|\bar u_{\st})\langle 0|\mC(\bar u_{\st}).
\een

\end{document}